\documentclass[nofootinbib,preprint,preprintnumbers,a4paper,11pt]{revtex4}
\usepackage{multirow}
\usepackage{textcomp}
\usepackage{amsmath,graphicx,color,epsfig}

\usepackage{footnote}
\usepackage{ulem}
\usepackage{booktabs}
\usepackage{array}
\usepackage{amssymb}
\usepackage{subfigure}
\usepackage{longtable}
\usepackage{verbatim}
\usepackage{amsfonts}
\usepackage{hyperref}
\usepackage{cancel}

\begin{document}

\title{$U$-spin conjugate $CP$ violation relations in bottom baryon decays}

\author{Bo-Nan Zhang$^{1}$}
\author{Di Wang$^{2}$}\email{wangdi@hunnu.edu.cn}

\address{%
$^1$College of Physics and Electronic Information, Inner Mongolia Normal University, Hohhot, 010022, Inner Mongolia, China\\
$^2$Department of Physics, Hunan Normal University, Changsha, 410081, China
}

\begin{abstract}
Motivated by the first observation of $CP$ violation in baryon decays, we study the relations for $CP$ asymmetries of bottom baryon decays.
It is found that all the $CP$ violation ratios of two $U$-spin conjugate decay channels satisfy a simple equation in the $U$-spin limit.
As examples, we present the $U$-spin conjugate channels in the four-body decays of bottom baryons.
The $CP$ violation ratios between them might provide predictions for unobserved channels and be tested in future experiments.

\end{abstract}

\maketitle


\section{Introduction}

Very recently, the LHCb Collaboration observed $CP$ violation in bottom baryon decays with the result of \cite{LHCb:2025ray}
\begin{align}
A_{CP}(\Lambda_b^0\to pK^-\pi^+\pi^-) = (2.45\pm 0.46\pm0.10)\%.
\end{align}
This discovery constitutes a milestone of particle physics since the $CP$ asymmetries are well established in meson systems \cite{LHCb:2019hro,Belle:2001zzw,Christenson:1964fg,BaBar:2001ags}, while the $CP$ violation in baryon decays had not been observed, except for a sole evidence in the three-body mode $\Lambda_b\to \Lambda K^+K^-$ \cite{LHCb:2024yzj}, until now.
$CP$ violation is a crucial element in interpreting the matter-antimatter asymmetry in the Universe \cite{Sakharov:1967dj} and serves as a window in searching for new physics.
Within the Standard Model, CP violation arises naturally through the Kobayashi-Maskawa (KM) mechanism \cite{Cabibbo:1963yz,Kobayashi:1973fv}, the study of which plays an important role in testing the Standard Model, and helps us to deepen our understanding the strong and electroweak interactions.

Flavor symmetries, including the $U-$,$V$-spin and isospin symmetries, provide model-independent tools to study heavy meson and baryon decays.
The $U$-spin symmetry, reflecting the symmetry between the two down-type $d$ and $s$ quarks, is the most powerful compared to the isospin and $V$-spin symmetry, since it connects the weak phase and $CP$ violation in weak decays.
The $U$-spin analysis for the $CP$ violation of bottom and charm hadron decays has been performed in literature such as \cite{Wang:2019dls,Wang:2024rwf,Schacht:2022kuj,Gavrilova:2022hbx,Dery:2021mll,Grossman:2018ptn,Bhattacharya:2015uua,Gronau:2013mda,Gronau:2000zy}.
The $U$-spin conjugation is defined by the following changes:
\begin{align}
s\leftrightarrow d,\qquad \overline s\leftrightarrow -\overline d.
\end{align}
Two decay channels are $U$-spin conjugate if they have $U$-spin conjugate initial and final states.
For example, the $\Lambda_b^0\to pK^-$ and $\Xi_b^0\to \Sigma^+\pi^-$ decays are two $U$-spin conjugate channels.
In literature such as Ref.~\cite{Wang:2024rwf}, a relation for the $CP$ violation ratio of the three-body bottom baryon decays is found.
In fact, this relation can be explained by a universal rule for two $U$-spin conjugate channels.
In this work, we illustrate this universal rule, and provide the relations between $CP$ asymmetries of the four-body decays of bottom baryons.

The rest of this paper is organized as follows. In Sec. \ref{BM}, we prove that the $CP$ violation ratio between two $U$-spin conjugate channels satisfies a universal law.
In Sec. \ref{MB}, we present the $U$-spin conjugate channels in the four-body decays of bottom baryons.
Sec. \ref{summary} is a brief summary of this work.

\section{Universal rule for $U$-spin conjugate channels}\label{BM}

In this section, we prove the universal rule about $CP$ asymmetries of two $U$-spin conjugate decay channels.
The effective Hamiltonian of bottom quark decay is given by \cite{Buchalla:1995vs}
\begin{align}\label{hsmb2}
 \mathcal H_{\rm eff}=&{\frac{G_F}{\sqrt 2} }
 \sum_{q=d,s}\left[V_{ub}^*V_{uq}\left(\sum_{i=1}^2C_i^u(\mu)O_i^u(\mu)\right) + V_{cb}^*V_{cq}\left(\sum_{i=1}^2C_i^c(\mu)O_i^c(\mu)\right)\right]\nonumber\\&
 -{\frac{G_F}{\sqrt 2}}\sum_{q=d,s}\left[V_{tb}V_{tq}^*\left(\sum_{i=3}^{10}C_i(\mu)O_i(\mu)
 +C_{7\gamma}(\mu)O_{7\gamma}(\mu)+C_{8g}(\mu)O_{8g}(\mu)\right)\right]+h.c..
 \end{align}
In the $SU(3)$ picture, the weak Hamiltonian of bottom decay can be written as \cite{Wang:2024ztg,Wang:2020gmn}
 \begin{equation}\label{h}
 \mathcal H_{\rm eff}= \sum_{i,j,k=1}^3 \{H^{(u)k}_{ij}\mathcal{O}_{ij}^{(u)k}+H^{(c)}_{i}\mathcal{O}_{i}^{(c)}
 +H^{(t)k}_{ij}\mathcal{O}_{ij}^{(t)k}\},
 \end{equation}
where $\mathcal{O}_{ij}^{(u)k}$, $\mathcal{O}_{i}^{(c)}$ and $\mathcal{O}_{ij}^{(t)k}$ denote the four-quark operator including the Fermi coupling constant $G_F$ and Wilson coefficients.
Superscripts $u$, $c$ and $t$ are used to distinguish the tree operators $\mathcal{O}_{1,2}^{(u)}$, $\mathcal{O}_{1,2}^{(c)}$ and penguin operators $\mathcal{O}_{3-10}$.
Indices $i$, $j$ and $k$ are flavor indices.
The matrix $H^{(u,c,t)}$ is the coefficient matrix.
According to Eq.~\eqref{h}, the non-zero CKM coefficients include:
\begin{align}\label{CKM}
  H^{(u)1}_{21} & = V_{ub}V_{ud}^*,\qquad H^{(u)1}_{31} = V_{ub}V_{us}^*, \qquad H^{(c)}_{2} = V_{cb}V_{cd}^*,\qquad H^{(c)}_{3} = V_{cb}V_{cs}^* \nonumber\\
 H^{(p)1}_{12} & = H^{(p)2}_{22}=H^{(p)3}_{32}=-V_{tb}V_{td}^*, \qquad H^{(p)1}_{13} = H^{(p)2}_{23}=H^{(p)3}_{33}= -V_{tb}V_{ts}^*.
\end{align}
The effective Hamiltonians for the $b \to d$ and $b \to s$ transitions change the $U$-spin and its third component as $|\frac{1}{2}, \frac{1}{2}\rangle$ and $|\frac{1}{2}, -\frac{1}{2}\rangle$, respectively.
They are $U$-spin conjugate since the up-type quarks are independent of $U$-spin.

If the $U$-spin representation of a particle is $|U,U_3\rangle$, the $U$-spin representation of its $U$-spin conjugate partner is $|U,-U_3\rangle$.
The $U$-spin amplitudes can be derived by coupling the weak Hamiltonian with the initial and final states,
\begin{align}
\langle f_1f_2&...f_n|\mathcal{H}_{b\to \mathcal{U}\overline {\mathcal{U}} q}|i\rangle=\langle f_1f_2...f_n|\mathcal{O}^{\mathcal{(U)}}_q|i\rangle\times H^{(\mathcal{U})}_q \nonumber\\&~~= \langle U^{f_1},U^{f_1}_3; U^{f_2},U^{f_2}_3;...U^{f_n},U^{f_n}_3|U^{\mathcal{H}},U^{\mathcal{H}}_3; U^{i},U^{i}_3\rangle\times V_{\mathcal{U}b}V_{\mathcal{U}q}^*,
\end{align}
where $\mathcal{U} = u,c,t$ and $q = d,s$.
According to the angular momentum coupling rules, the Clebsch-Gordan (CG) coefficients satisfy the following relation \cite{Varshalovich:1988ifq}:
\begin{align}\label{p}
 \langle &j_1m_1j_2m_2...j_\alpha m_\alpha|j_{\alpha+1}m_{\alpha+1}j_{\alpha+2}m_{\alpha+2}\rangle
 \nonumber\\&~~~~=(-1)^{j_1+j_2+...+j_\alpha-j_{\alpha+1}-j_{\alpha+2}}\langle j_1-m_1j_2-m_2...j_\alpha-m_\alpha|j_{\alpha+1}-m_{\alpha+1}j_{\alpha+2}-m_{\alpha+2}\rangle.
\end{align}
If the decay amplitude of a $\Delta S = 0$ mode $i\to f$ is written as
\begin{align}\label{a2}
 \mathcal{A}(i\to f)=V_{ub}V_{ud}^*\mathcal{A}^u+ V_{cb}V_{cd}^*\mathcal{A}^c+V_{tb}V_{td}^*\mathcal{A}^t,
\end{align}
the decay amplitude of the $U$-spin conjugate $\Delta S = -1$ mode $i^\prime\to f^\prime$ is
\begin{align}\label{a3}
 \mathcal{A}(i^\prime\to f^\prime)=\pm (V_{ub}V_{us}^*\mathcal{A}^u+ V_{cb}V_{cs}^*\mathcal{A}^c+V_{tb}V_{ts}^*\mathcal{A}^t).
\end{align}
The decay amplitudes of two $U$-spin conjugate modes are related by the following interchanges of the CKM matrix elements:
\begin{align}\label{a1}
  V_{ub}V_{ud}^* \,\leftrightarrow \,V_{ub}V_{us}^*, \qquad  V_{cb}V_{cd}^*\, \leftrightarrow  \,V_{cb}V_{cs}^*, \qquad  V_{tb}V_{td}^* \,\leftrightarrow \, V_{tb}V_{ts}^*,
\end{align}
together with a possible overall minus sign.

As examples, we derive the $U$-spin amplitudes for the $\Lambda_b^0\to pK^-\pi^+\pi^-$ and $\Xi _b^0\to\Sigma^+ \pi^-K^+K^-$ decays.
The pairs $(\Lambda_b^0,\Xi_b^0)$, $(p,\Sigma^+)$, $(K^+,\pi^+)$, and $(\pi^-,K^-)$ form four $U$-spin doublets.
According to the coupling rule in Eq.~\eqref{p}, the amplitudes induced by the $b\to u\overline us$ and $b\to u\overline ud$ transitions have following relation,
\begin{align}\label{x}
 \frac{\langle pK^-\pi^+\pi^-|\mathcal{H}_{b\to u\overline us}|\Lambda_b^0\rangle}{\langle \Sigma ^+ \pi^-K^+K^-|\mathcal{H}_{b\to u\overline ud}|\Xi_b^0\rangle} = -\frac{V_{ub}\,V_{us}^*}{V_{ub}V_{ud}^*},
\end{align}
in which the minus sign arises from $(-1)^{\frac{1}{2}\times 4-\frac{1}{2}\times 2} = -1$.
The $U$-spin amplitude for the $\Lambda_b^0\to pK^-\pi^+\pi^-$ decay induced by $b\to u\overline us$ transition is calculated to be
\begin{align}\label{b1}
&\qquad\,\langle pK^-\pi^+\pi^-|\mathcal{H}_{b\to u\overline us}|\Lambda_b^0\rangle\nonumber\\&= \langle \frac{1}{2},\frac{1}{2}; \frac{1}{2},-\frac{1}{2};\frac{1}{2},-\frac{1}{2}; \frac{1}{2},\frac{1}{2}|\frac{1}{2},-\frac{1}{2}; \frac{1}{2},\frac{1}{2}\rangle\times V_{ub}V_{us}^*\nonumber\\&=(\frac{1}{\sqrt{6}}\langle 2,0|-\frac{1}{\sqrt{6}}\langle 1,0|^{(1)}+ \frac{1}{2\sqrt{3}}\langle 1,0|^{(2)}-\frac{1}{2\sqrt{3}}\langle 0,0|^{(1)}+\frac{1}{2}\langle 1,0|^{(3)}-\frac{1}{2}\langle 0,0|^{(2)})\nonumber\\&\qquad\qquad (\frac{1}{\sqrt{2}}|1,0\rangle-\frac{1}{\sqrt{2}}| 0,0\rangle)\times V_{ub}V_{us}^*\nonumber\\&=(-\frac{1}{2\sqrt{3}}\,\mathcal{A}^{(1)}_{1}
+\frac{1}{2\sqrt{6}}\,\mathcal{A}^{(2)}_{1}
+\frac{1}{2\sqrt{2}}\,\mathcal{A}^{(3)}_{1}
+\frac{1}{2\sqrt{6}}\,\mathcal{A}^{(1)}_{0}
+\frac{1}{2\sqrt{2}}\,\mathcal{A}^{(2)}_{0})\times V_{ub}V_{us}^*.
\end{align}
And the $U$-spin amplitude for the $\Xi _b^0\to\Sigma^+ \pi^-K^+K^- $ mode induced by $b\to u\overline ud$ transition is calculated to be
\begin{align}\label{b2}
&\qquad\,\langle \Sigma ^+ \pi^-K^+K^-|\mathcal{H}_{b\to u\overline ud}|\Xi_b^0\rangle\nonumber\\&= \langle \frac{1}{2},-\frac{1}{2}; \frac{1}{2},\frac{1}{2};\frac{1}{2},-\frac{1}{2}; \frac{1}{2},\frac{1}{2}|\frac{1}{2},\frac{1}{2}; \frac{1}{2},-\frac{1}{2}\rangle\times V_{ub}V_{ud}^*\nonumber\\&=(\frac{1}{\sqrt{6}}\langle 2,0|+\frac{1}{\sqrt{6}}\langle 1,0|^{(1)}- \frac{1}{2\sqrt{3}}\langle 1,0|^{(2)}-\frac{1}{2\sqrt{3}}\langle 0,0|^{(1)}-\frac{1}{2}\langle 1,0|^{(3)}-\frac{1}{2}\langle 0,0|^{(2)})\nonumber\\&\qquad\qquad (\frac{1}{\sqrt{2}}|1,0\rangle-\frac{1}{\sqrt{2}}| 0,0\rangle)\times V_{ub}V_{ud}^*\nonumber\\&=(\frac{1}{2\sqrt{3}}\,\mathcal{A}^{(1)}_{1}
-\frac{1}{2\sqrt{6}}\,\mathcal{A}^{(2)}_{1}
-\frac{1}{2\sqrt{2}}\,\mathcal{A}^{(3)}_{1}
-\frac{1}{2\sqrt{6}}\,\mathcal{A}^{(1)}_{0}
-\frac{1}{2\sqrt{2}}\,\mathcal{A}^{(2)}_{0})\times V_{ub}V_{ud}^*.
\end{align}
These two $U$-spin amplitudes are consistent with Eq.~\eqref{x}.
Together with the amplitudes induced by the $b\to c\overline cd/s$ transition and penguin contributions, the relation given by Eqs.~\eqref{a2} and \eqref{a3} is obtained.

The unitarity of the CKM matrix implies that
\begin{align}\label{a7}
 V_{ub}V_{ud}^*+ V_{cb}V_{cd}^*+V_{tb}V_{td}^*=0.
\end{align}
Then the decay amplitude of $i\to f$ mode is
\begin{align}
 \mathcal{A}(i\to f)&=V_{ub}V_{ud}^*\mathcal{A}^u+ V_{cb}V_{cd}^*\mathcal{A}^c-(V_{ub}V_{ud}^*+V_{cb}V_{cd}^*)\mathcal{A}^t
 \nonumber\\&~~=V_{ub}V_{ud}^*(\mathcal{A}^u-\mathcal{A}^t)
 +V_{cb}V_{cd}^*(\mathcal{A}^c-\mathcal{A}^t) = V_{ub}V_{ud}^*\mathcal{A}^{\prime u}
 +V_{cb}V_{cd}^*\mathcal{A}^{\prime c}.
\end{align}
Similarly, the decay amplitude of $i^\prime\to f^\prime$ mode is
\begin{align}
 \mathcal{A}(i^\prime\to f^\prime)=\pm(V_{ub}V_{us}^*\mathcal{A}^{\prime u}+ V_{cb}V_{cs}^*\mathcal{A}^{\prime c}).
\end{align}
The direct $CP$ asymmetry of $i\to f$ decay is defined as
\begin{align}\label{a9}
  A_{CP}^{\rm dir}(i\to f) =\frac{|\mathcal{A}(i\to f)|^2-\mathcal{A}(\overline i\to \overline f)|^2}{|\mathcal{A}(i\to f)|^2+\mathcal{A}(\overline i\to \overline f)|^2},
\end{align}
where
\begin{align}\label{a8}
  \mathcal{A}(\overline i\to \overline f) = V^*_{ub}V_{ud}\mathcal{A}^{\prime u}
 +V^*_{cb}V_{cd}\mathcal{A}^{\prime c}.
\end{align}
Substituting Eq.~\eqref{a7} and Eq.~\eqref{a8} into Eq.~\eqref{a9}, the direct $CP$ violation $A_{CP}^{\rm dir}(i\to f)$ is derived to be
\begin{align}
  A_{CP}^{\rm dir}(i\to f) \simeq \frac{2\,\mathcal{I}m[V_{ub}V_{ud}^*V_{cb}^*V_{cd}]\,\mathcal{I}m[\mathcal{A}^{\prime u}(\mathcal{A}^{\prime c})^*]} {|\mathcal{A}(i\to f)|^2},
\end{align}
where the approximation $|\mathcal{A}(i\to f)|^2+|\mathcal{A}(\overline i\to \overline f)|^2\sim 2|\mathcal{A}(i\to f)|^2$ is used.
For the $i^\prime\to f^\prime$ decay,
the direct $CP$ violation is
\begin{align}
  A_{CP}^{\rm dir}(i^\prime\to f^\prime) \simeq \frac{2\,\mathcal{I}m[V_{ub}V_{us}^*V_{cb}^*V_{cs}]\,\mathcal{I}m[\mathcal{A}^{\prime u}(\mathcal{A}^{\prime c})^*]} {|\mathcal{A}(i^\prime\to f^\prime)|^2},
\end{align}
According to the unitarity of the CKM matrix, we have $\mathcal{I}m[V_{ub}V_{ud}^*V_{cb}^*V_{cd}]=-\mathcal{I}m[V_{ub}V_{us}^*V_{cb}^*V_{cs}]$.
Then the relation between the $CP$ asymmetries of two $U$-spin conjugate modes is derived to be
\begin{align}\label{a}
 A_{CP}^{\rm dir}(i^\prime\to f^\prime)\simeq -A_{CP}^{\rm dir}(i\to f)\cdot \frac{|\mathcal{A}(i\to f)|^2}{|\mathcal{A}( i^\prime\to  f^\prime)|^2}=-A_{CP}^{\rm dir}(i\to f)\cdot \frac{\mathcal{B}r(i\to f)\cdot\tau(i^\prime)}{\mathcal{B}r(i^\prime\to f^\prime)\cdot\tau(i)}.
\end{align}

\section{Four-body bottom baryon decays}\label{MB}
\begin{table*}
\caption{$U$-spin conjugate channels in the charmless four-body decays of bottom baryons.}\label{ta2}
 \small
\begin{tabular}{|c|c||c|c|}
\hline\hline
 \quad\quad $\Delta S = -1$ mode\qquad\qquad & \quad\quad$\Delta S = 0$ mode \qquad\qquad& \quad\quad$\Delta S = -1$ mode\qquad \qquad& \quad\quad$\Delta S = 0$ mode\qquad\qquad\\\hline
 $\Xi _b^-\to p K^- \overline K^0 \pi^-$ & $\Xi _b^-\to\Sigma ^+ \pi ^- K^0 K^-$ &$\Xi _b^-\to\Xi^- K^+ \overline K^0 \pi^-$ & $\Xi _b^-\to\Sigma ^- \pi ^+ K^0 K^-$\\\hline
 $\Xi _b^-\to \Xi^- K^- \pi^+  K^0$ & $\Xi _b^-\to \Sigma ^- \pi ^- K^+ \overline K^0$ & $\Xi _b^-\to \Sigma^+ K^- K^- K^0$& $\Xi _b^-\to p \pi ^- \pi ^- \overline{K}^0$\\\hline
 $\Xi_b^- \to \Sigma^+\overline K^0 \pi^- \pi^-$ & $\Xi _b^- \to p K^0 K^- K^-$ &$\Xi _b^-\to \Xi^0 K^+ K^- \pi^-$ & $\Xi _b^-\to n \pi ^+ \pi ^- K^-$\\\hline
 $\Xi _b^-\to \Xi^0 \pi^+ \pi^- \pi^-$ & $\Xi _b^-\to n K^+ K^- K^-$ &$\Xi _b^-\to \Xi^0 \overline K^0 K^0 \pi^-$ & $\Xi _b^-\to n K^0 \overline{K}^0 K^-$\\\hline
$\Xi _b^-\to n K^- K^- \pi^+$  & $\Xi _b^-\to\Xi ^0 \pi ^- \pi ^- K^+$ &$\Xi _b^-\to n \overline K^0 \overline K^0 \pi^-$ & $\Xi _b^-\to\Xi ^0 K^0 K^0 K^-$\\\hline
 $\Xi _b^-\to\Sigma^- K^+ K^- \overline K^0$ & $\Xi _b^-\to\Xi ^- \pi ^+ \pi ^- K^0$ & $\Xi _b^-\to\Sigma^- \pi^+ \overline K^0 \pi^-$& $\Xi _b^-\to\Xi ^- K^+ K^0 K^-$\\\hline
 $\Xi _b^-\to\Sigma^- \overline K^0 \overline K^0 K^0$ & $\Xi _b^-\to\Xi ^- K^0 K^0 \overline K^0$ &$\Lambda _b^0\to p K^+ K^- K^-$ & $\Xi _b^0\to\Sigma ^+ \pi ^+ \pi ^- \pi ^-$\\\hline
 $\Lambda_b^0\to p K^- \pi^+ \pi^-$ & $\Xi _b^0\to\Sigma ^+ \pi ^- K^+ K^-$ & $\Lambda_b^0\to p K^- \overline K^0 K^0$& $\Xi _b^0\to\Sigma ^+ \pi ^- K^0 \overline K^0$\\\hline
 $\Lambda_b^0\to \Xi^- K^+ K^+K^-   $ & $\Xi _b^0\to \Sigma ^-\pi ^+ \pi ^+\pi ^-   $ & $\Lambda_b^0 \to \Xi^- K^+ \pi^- \pi^+ $& $\Xi _b^0 \to \Sigma ^-\pi ^+ K^- K^+ $\\\hline
 $\Lambda_b^0\to \Xi^- K^+ \overline K^0   \bar{K}^0 $ & $\Xi _b^0\to \Sigma ^- \pi ^+ K^0   \overline{K}^0 $ & $\Lambda_b^0\to \Sigma^+ K^+ K^- \pi^-$& $\Xi _b^0\to p \pi ^+ \pi ^- K^-$\\\hline
 $\Lambda _b^0\to \Sigma^+ \pi^+ \pi^- \pi^-$ & $\Xi _b^0\to p K^+ K^- K^-$ & $\Lambda_b^0\to \Sigma^+ \overline K^0 K^0 \pi^-$ & $\Xi _b^0\to p K^0 \overline{K}^0 K^-$\\\hline
 $\Lambda _b^0\to \Xi^0 K^+ K^- K^0$ & $\Xi _b^0\to n \pi ^+ \pi ^- \overline{K}^0$ & $\Lambda_b^0 \to \Xi^0 \pi^- \pi^+ K^0$& $\Xi _b^0 \to n K^- K^+ \overline{K}^0$\\\hline
 $\Lambda_b^0\to \Xi^0 \overline K^0 K^0 K^0$ & $\Xi _b^0\to n K^0 \overline{K}^0 \overline{K}^0$ &$\Lambda_b^0\to n  \overline K^0 K^- K^+$ & $\Xi _b^0\to \Xi ^0  K^0 \pi ^- \pi ^+$\\\hline
$\Lambda_b^0\to n \pi^+ \overline K^0 \pi^-$  & $\Xi _b^0\to\Xi ^0 K^+ K^0 K^-$ & $\Lambda_b^0\to n \overline K^0 \overline K^0 K^0$& $\Xi _b^0\to\Xi ^0 K^0 K^0 \overline {K}^0$\\\hline
 $\Lambda_b^0\to\Sigma^- K^+ K^- \pi^+$ & $\Xi _b^0\to\Xi ^- \pi ^+ \pi ^- K^+$ & $\Lambda_b^0\to\Sigma^- \pi^+ \pi^+ \pi^-$& $\Xi _b^0\to\Xi ^- K^+ K^+ K^-$\\\hline
 $\Lambda_b^0\to\Sigma^- \pi^+ \overline K^0 K^0$ & $\Xi _b^0\to\Xi ^- K^+ K^0 \overline {K}^0$ & $\Xi_b^0\to  \Xi^+\pi^+ K^- K^- $& $\Lambda _b^0\to  \Sigma ^+K^+ \pi ^- \pi ^- $\\\hline
 $\Xi_b^0\to p \overline K^0 \overline K^0 \pi^-$ & $\Lambda _b^0\to\Sigma ^+ K^0 K^0 K^-$ & $\Xi_b^0 \to \Xi^-\pi^+ K^- K^+ $& $\Lambda _b^0 \to \Sigma ^-K^+ \pi ^- \pi ^+ $\\\hline
$\Xi_b^0\to\Xi^- \pi^+ \pi^+ \pi^-$  & $\Lambda _b^0\to\Sigma ^- K^+ K^+ K^-$ & $\Xi_b^0\to\Xi^- \pi^+\overline K^0 K^0$& $\Lambda _b^0\to\Sigma ^- K^+ K^0 \overline{K}^0$\\\hline
 $\Xi_b^0\to \Sigma^+ K^+ K^- K^-$ & $\Lambda _b^0\to p \pi ^+ \pi ^- \pi ^-$ &$\Xi_b^0\to \Sigma^+ K^- \pi^+ \pi^-$ & $\Lambda _b^0\to p \pi ^- K^+ K^-$\\\hline
 $\Xi_b^0\to \Sigma^+ K^- \overline K^0 K^0$ & $\Lambda _b^0\to p \pi ^- K^0 \overline{K}^0$ & $\Xi_b^0\to \Xi^0 K^+ \overline K^0 \pi^-$& $\Lambda _b^0\to n \pi ^+ K^0 K^-$\\\hline
 $\Xi_b^0\to \Xi^0 K^- \pi^+ K^0$ & $\Lambda _b^0\to n \pi ^- K^+ \overline{K}^0$ &$\Xi_b^0\to n K^- \pi^+ \overline K^0$ & $\Lambda _b^0\to\Xi ^0 \pi ^- K^+ K^0$\\\hline
 $\Xi_b^0\to\Sigma^- K^+ \overline K^0 \overline K^0$ & $\Lambda _b^0\to\Xi ^- \pi ^+ K^0 K^0$ & $\Xi_b^0\to\Sigma^- K^- \pi^+ \pi^+$& $\Lambda _b^0\to\Xi ^- \pi ^- K^+ K^+$\\\hline
  \hline
\end{tabular}
\end{table*}

The $U$-spin conjugation transformations for bottom baryons, pseudoscalar mesons, and octet and decuplet baryons are given by
\begin{align}
\Lambda^0_b \leftrightarrow \Xi^0_b, \quad \Xi^-_b \leftrightarrow \Xi^-_b,
\end{align}
\begin{align}
K^+ \leftrightarrow \pi^+, \quad K^- \leftrightarrow \pi^-,  \quad K^0\leftrightarrow \overline K^0,
\end{align}
\begin{align}
p\leftrightarrow \Sigma^+,\quad n\leftrightarrow \Xi^0,\quad \Sigma^-\leftrightarrow \Xi^-,
\end{align}
\begin{align}
\Delta^0\leftrightarrow \Xi^{*0},\quad \Sigma^{*+}\leftrightarrow \Delta^+,\quad \Sigma^{*0}\leftrightarrow \Sigma^{*0}, \quad \Delta^{++}\leftrightarrow \Delta^{++},\quad \Xi^{*-}\leftrightarrow \Sigma^{*-},\quad \Delta^{-}\leftrightarrow \Omega^{-}.
\end{align}
For the excited states, the $U$-spin conjugation transformation is similar to the ground states.
According to these $U$-spin conjugation transformation rules, we can find many decay channel pairs whose $CP$ violation ratios satisfy Eq.~\eqref{a}.
For examples, the $U$-spin conjugate channels in the charmless four-body decays of bottom baryons are listed in Table.~\ref{ta2}.

Similarly to bottom hadron decays, the $CP$ asymmetries of two $U$-spin conjugate charmed hadron decay channels are related by a simple rule.
For a singly cabibbo-suppressed charmed hadron decay mode, both the $c\to d\overline du$ and $c\to s\overline su$ transitions contribute.
The $c\to d\overline du$ and $c\to s\overline su$ transitions are $U$-spin conjugate.
If decay amplitude of $i\to f$ mode is written as
\begin{align}\label{a10}
 \mathcal{A}(i\to f)& = V_{cd}^*V_{ud}\mathcal{A}_1
 +V_{cs}^*V_{us}\mathcal{A}_2,
\end{align}
the decay amplitude of its $U$-spin conjugate mode $i^\prime\to f^\prime$ is
\begin{align}\label{a11}
 \mathcal{A}(i^\prime\to f^\prime)= \pm(V_{cs}^*V_{us}\mathcal{A}_1
 +V_{cd}^*V_{ud}\mathcal{A}_2).
\end{align}
Substituting Eq.~\eqref{a10} and the corresponding $\mathcal{A}(\overline i\to \overline f)$ into Eq.~\eqref{a9}, the direct $CP$ violation for the $i\to f$ decay is derived to be
\begin{align}
  A_{CP}^{\rm dir}(i\to f) \simeq \frac{2\,\mathcal{I}m[V_{cd}^*V_{ud}V_{cs}V_{us}^*]\,\mathcal{I}m[\mathcal{A}_1\mathcal{A}_2^*]} {|\mathcal{A}(i\to f)|^2}.
\end{align}
For the $CP$ violation of $i^\prime\to f^\prime$ decay, $\mathcal{I}m[V_{cd}^*V_{ud}V_{cs}V_{us}^*]$ is replaced by $\mathcal{I}m[V_{cs}^*V_{us}V_{cd}V_{ud}^*] = -\mathcal{I}m[V_{cd}^*V_{ud}V_{cs}V_{us}^*]$.
Unlike bottom hadron decays, we have $|\mathcal{A}(i\to f)|^2 \simeq |\mathcal{A}( i^\prime\to  f^\prime)|^2$ for charmed hadron decays due to the approximation $V^*_{cd}V_{ud}\simeq - V^*_{cs}V_{us}$.
Then Eq.~\eqref{a} is degenerated into
\begin{align}\label{ax}
 A_{CP}^{\rm dir}(i^\prime\to f^\prime)\simeq -A_{CP}^{\rm dir}(i\to f).
\end{align}
More detailed analysis of charmed hadron decays can be found in Ref.~\cite{Wang:2019dls}.

The relation given by Eq.~\eqref{a} is derived in the $U$-spin limit.
When considering $U$-spin breaking, Eq.~\eqref{a} is no longer valid.
The $U$-spin breaking is naively estimated to be $m_s/\Lambda_{\rm QCD}\sim 30\%$. Due to the lack of experimental data, the $U$-spin breaking in bottom baryon decays cannot be extracted so far.
However, we can estimate the $U$-spin breaking in bottom baryon decays from the data of bottom meson decays indirectly.
The $B^0\to K^+\pi^-$ and $B^0_s\to K^-\pi^+$ modes are $U$-spin conjugate.
In the $U$-spin limit, their $CP$ asymmetries satisfy Eq.~\eqref{a}.
The lifetime ratio between $B^0$ and $B^0_s$ is $\tau_{B^0}/\tau_{B^0_s} \approx 1.0$ \cite{PDG}. Thus, we have
\begin{align}
  \frac{A_{CP}^{\rm dir}(B^0\to K^+\pi^-)}{A_{CP}^{\rm dir}(B^0_s\to K^-\pi^+)}\simeq-\frac{\mathcal{B}r(B^0_s\to K^-\pi^+)}{\mathcal{B}r(B^0\to K^+\pi^-)}.
\end{align}
The $CP$ asymmetries of these two decay modes are given by \cite{PDG}
\begin{align}
  A_{CP}^{\rm dir}(B^0\to K^+\pi^-)& = -0.0836\pm 0.0032,\nonumber\\ A_{CP}^{\rm dir}(B^0_s\to K^-\pi^+) &=0.225\pm 0.012.
\end{align}
The branching fractions are \cite{PDG}
\begin{align}
  \mathcal{B}r(B^0\to K^+\pi^-) &= (2.00\pm 0.04)\times 10^{-5},\nonumber\\ \mathcal{B}r(B^0_s\to K^-\pi^+) &=(0.59\pm 0.07)\times 10^{-5}.
\end{align}
The ratios of asymmetries and branching fractions are
\begin{align}
  \frac{A_{CP}^{\rm dir}(B^0\to K^+\pi^-)}{A_{CP}^{\rm dir}(B^0_s\to K^-\pi^+)}=-(37.2\pm2.4)\%,\qquad \frac{\mathcal{B}r(B^0_s\to K^-\pi^+)}{\mathcal{B}r(B^0\to K^+\pi^-)}=(29.5\pm3.6)\%.
\end{align}
It can be found the $U$-spin breaking in these two decay modes is consistent with the natively estimation.
Thus, we predict the uncertainty in Eq.~\eqref{a} induced by the $U$-spin breaking in bottom hadron decays is around $30\%$ too.
This level of uncertainty is acceptable when Eq.~\eqref{a} is used for theoretical predictions.

Eq.~\eqref{a} indicates that the $CP$ asymmetries in $U$-spin conjugate bottom baryon decay modes have opposite sign, just like the $CP$ asymmetries in the $B^0\to K^+\pi^-$ and $B^0_s\to K^-\pi^+$ decays.
We can use Table. \ref{ta2} and Eq.~\eqref{a} to construct some observables in which two $CP$ asymmetries are constructive.
For example, the $CP$ asymmetry difference
\begin{align}
 \Delta A_{CP}^{\rm }&= A_{CP}(\Lambda_{b}^{0}\to pK^{-}\pi^+\pi^-) -  A_{CP}(\Xi_{b}^{0}\to \Sigma^+\pi^{-}K^+K^-)
\end{align}
is constructive.
The $CP$ violation relations between two $U$-spin conjugate channels could provide predictions for unobserved channels and be tested in future experiments.

The topic of this work has also studied in Ref.~\cite{Wang:2024rwf} and in Refs.~\cite{He:2025msg,Chen:2025drl,Roy:2025nao}, which appeared before and after this work, respectively.
In both Ref.~\cite{Roy:2025nao} and this work, the $U$-spin amplitudes for four-body decays are derived through direct $U$-spin coupling calculations.
This approach allows for different choices of $U$-spin coupling order.
Different coupling orders correspond to different representations in quantum mechanics, which are related by unitary transformation matrices.
To compare the relations between two $U$-spin conjugate channels, it is convenient to express the $U$-spin amplitudes using the one-by-one $U$-spin conjugation order, as demonstrated in Eqs.~\eqref{b1} and \eqref{b2}.
In Ref.~\cite{Roy:2025nao}, some of the $U$-spin amplitudes for $U$-spin conjugate channels are not written in this order, resulting in the not manifest $U$-spin relations.

In Refs.~\cite{Wang:2024rwf,He:2025msg,Chen:2025drl}, the $U$-spin amplitudes for three- or four-body decays are constructed via tensor contractions.
In the $SU(3)_F$ limit, pseudoscalar mesons in the final state are treated as identical particles.
According to the generalized identity principle, the Hamiltonian is symmetrized if the two mesons are in a relatively even angular momentum state or antisymmetrized if they are in a relatively odd angular momentum state \cite{Savage:1989ub,Savage:1989qr}.
The decay amplitudes constructed via tensor contractions depend on the momentum configuration of the final states.
In multi-body decays, various intermediate resonances contribute, leading to diverse momentum configurations.
Thus, the decay amplitudes for multi-body decays and the $CP$ violation relations beyond Eq.~\eqref{a} merit further investigation.

\section{Summary}\label{summary}

In this work, we studied the relations for $CP$ asymmetries of bottom baryon decays.
It is found that there is a universal rule for the $CP$ violation ratio between two $U$-spin conjugate modes in the $U$-spin limit.
As examples, the $U$-spin conjugate channels in charmless four-body decays, which satisfy the universal rule, are presented.
The $CP$ violation relations between them could provide predictions for unobserved channels and enable testing of $U$-spin symmetry in future measurements.

\begin{acknowledgements}

We thank Zhi-Heng Wang for providing theoretical support about angular momentum coupling.
This work was supported in part by the National Natural Science Foundation of China under Grants No. 12105099.

\end{acknowledgements}

\end{document}